\documentstyle[12pt]{article}

\newcommand{\eq}{\begin{equation}}
\newcommand{\eqn}{\begin{displaymath}}
\newcommand{\en}{\end{equation}}
\newcommand{\enn}{\end{displaymath}}

\def\pnot{\mbox{${\not{\hbox{\kern-3.0pt$p$}}}$}}
\def\qnot{\mbox{${\not{\hbox{\kern-2.0pt$q$}}}$}}
\def\enot{\mbox{${\not{\hbox{\kern-2.0pt$e$}}}$}}
\def\knot{\mbox{${\not{\hbox{\kern-2.0pt$k$}}}$}}

\def\fun#1#2{\lower3.6pt\vbox{\baselineskip0pt\lineskip.9pt\ialign
{$\mathsurround=0pt#1\hfil##\hfil$\crcr#2\crcr\sim\crcr}}}

\begin{document}


\begin{titlepage}
\hskip 12cm \vbox{\hbox{BUDKERINP/94-46}\hbox{DF-UNICAL/94-14}\hbox{May 1994}}
\vskip 0.6cm
\centerline{\bf QUARK CONTRIBUTION}
\centerline{\bf TO THE REGGEON-REGGEON-GLUON VERTEX IN QCD$^{~\diamond}$}
\vskip 1.0cm
\centerline{  V.Fadin$^{\dagger}$}
\vskip .5cm
\centerline{\sl Budker Institute for Nuclear Physics}
\centerline{\sl and Novosibirsk State University, 630090 Novosibirsk,
Russia}
\vskip .5cm
\centerline{  R.Fiore$^{\ddagger}$, A.Quartarolo$^{\ddagger}$}
\vskip .5cm
\centerline{\sl  Dipartimento di Fisica, Universit\`a della Calabria}
\centerline{\sl Istituto Nazionale di Fisica Nucleare, Gruppo collegato di
Cosenza}
\centerline{\sl Arcavacata di Rende, I-87036 Cosenza, Italy}
\vskip 1cm
\begin{abstract}
The quark loop contribution to the reggeon-reggeon-gluon vertex is calculated
in QCD, where the reggeon is the reggeized gluon. Compared with the vertex in
the Born approximation, this contribution exhibits a new spin structure as
well as the gluon loop one. A remarkable but not complete cancellation
between gluon and quark contributions to this new spin structure takes place
for the case of three massless quark flavours.
\end{abstract}
\vskip .5cm
\hrule
\vskip.3cm
\noindent

\noindent
$^{\diamond}${\it Work supported in part by the Ministero italiano
dell'Universit\`a e della Ricerca Scientifica e Tecnologica and in part by
the EEC Programme ``Human Capital and Mobility", Network ``Physics at High
Energy Colliders", contract CHRX-CT93-0537 (DG 12 COMA)}
\vfill
$ \begin{array}{ll}
^{\dagger}\mbox{{\it email address:}} &
 \mbox{FADIN~@INP.NSK.SU}\\
\end{array}
$

$ \begin{array}{ll}
^{\ddagger}\mbox{{\it email address:}} &

   \mbox{39022::FIORE, QUARTAROLO}\\
 & \mbox{FIORE, QUARTAROLO~@FIS.UNICAL.IT}
\end{array}
$
\vfill
\end{titlepage}
\eject
\textheight 210mm
\topmargin 2mm
\baselineskip=24pt
{\bf 1. Introduction}

Semihard processes play a more and more important role at present colliders
and, presumably, this role will increase in the future. Since the typical
virtuality $Q^2$ for these processes is large enough to ensure a smallness
of the strong coupling constant $\alpha_s(Q^2)$, perturbation theory can be
applied to calculate parton distributions and cross sections. On the other
hand, the
energy $\sqrt{s}$ of colliding particles is
sufficiently high to make the ratio ${x={{Q^2}\over s}}$ so small that the
problem of summing up logarithmic terms ${{\alpha_s}^n}{(ln{1\over x})^m}$
arises. In the leading logarithmic approximation (LLA), which means $n=m$,
this problem was solved many years ago~\cite{FKL}.
\vskip.3cm
Unfortunately in the LLA unitarity is violated and the pomeron singularity
in the $j$-plane lies on the r.h.s. of unity, yielding a strong
power increase of structure functions in the small $x$ region. Therefore the
evaluation of next-to-leading terms is necessary for defining the region in
which the LLA can be applied, as well as for fixing the pomeron intercept.
\vskip.3cm
The calculation of radiative corrections to the LLA was started
by L.N. Lipatov and one of the authors (V.F.) in ref.~\cite{LF}. The
calculation program is based on the gluon reggeization property that was
proved in the LLA and that simplified essentially the derivation of the
equation for the pomeron. This equation is constructed in terms of the
reggeon (reggeized gluon) trajectory
\eq
j(t) = 1+\omega(t)~,
\label{z1}
\en
and the reggeon-reggeon-gluon (RRG) vertex. In the leading order, in the
case of the SU(N) gauge group ($N=3$ for QCD), one finds
\eq
\omega(t)=\omega^{(1)}(t)={g^2t\over {(2\pi)}^{3+\varepsilon}}{N \over 2}
\int{d^{2+\varepsilon}k_\perp\over k_\perp^2{(q-k)}_{\perp}^2}~,
\label{z2}
\en
where g is the coupling constant of the gauge theory, $q$ is the momentum
transfer and $t = q_\perp^2$. The integration is performed over the
two-dimensional momentum orthogonal to the initial particle momentum plane,
and dimensional regularization of Feynman integrals is used:
\eq
{d^2k\over {{(2\pi)}^2}}~~\rightarrow ~~{d^{2+\varepsilon}k\over {{(2\pi)}^
{2+\varepsilon}}},~~~~~~~\varepsilon = D-4~,
\label{z3}
\en
where $D$ is the space-time dimension $(D=4$ for the physical case).
\vskip.3cm
In the LLA the only essential kinematics is the multi-Regge kinematics (see
Fig. 1):
\eqn
s={(p_A+p_B)}^2 \gg s_i \gg {|t_i|}~,~~~~i=1,\cdots ,n+1~,
\enn
\eqn
s_1\cdot s_2\cdots s_{n+1}=s\prod^n_{j=1}\vec k_{j\perp}^{\thinspace 2}~,
{}~~~~s_i=(k_{i-1}+k_i)^2~,~~~~t_i=q^2_i~,
\enn
\eq
q_{i}=q_{i-1}-k_{i-1}~,~~~~q_0\equiv p_A~,~~~~k_0\equiv p_{A'}~~~~k_{n+1}
\equiv p_{B'}~.
\label{z4}
\en
Here $p_A$, $p_B$ and $p_{A'}$, $p_{B'}$ are the momenta of the
colliding and scattered particles respectively, $k_1$,....,$k_n$
are the momenta of the produced gluons and $k_{1\perp}$,...,$k_{n\perp}$ are
their transverse components
($k_{i\perp}^{\thinspace 2} = - \vec k_{i\perp}^{\thinspace 2}$).
Due to the gluon reggeization, the multi-gluon
production amplitude has a simple multi-Regge form in the kinematics
(\ref{z4})~\cite{FKL}:
\eq
{\cal A}^{A'G_1\ldots G_nB'}_{AB} = 2s\Gamma^{i_1}_{A'A}{s^{\omega(t_1)}_1
\over {t_1}}\gamma^{G_1}_{i_1i_2}(q_1,q_2){s^{\omega(t_2)}_2\over
{t_2}}\gamma^{G_2}_{i_2i_3}(q_2,q_3)\cdot\cdot\cdot\cdot{s^{\omega(t_
{n+1})}_{n+1}\over{t_{n+1}}}\Gamma^{i_{n+1}}_{B'B}~,
\label{z5}
\en
where $\Gamma^i_{A'A}$ and $\gamma^G_{ij}(q_1,q_2)$ are the
particle-particle-reggeon (PPR) and the RRG vertices respectively.
In the LLA the helicity of each of the colliding particles is conserved and
in the helicity basis the former vertices take the form
\eq
\Gamma^{i}_{A'A}=\Gamma^{(0)i}_{A'A} = gT^i_{A'A}\delta_{{\lambda_{A'}}
{\lambda_A}}~,
\label{z6}
\en
where $\lambda_A$ is the helicity of the particle $A$ and $T^i_{A'A}$
stands for the matrix elements of the colour group generators in the
corresponding representation (i.e. adjoint for gluons and fundamental,
$T^i= t^i = {\lambda^i\over 2}$, for quarks). The RRG vertex in turn is
written as
\eq
\gamma^G_{ij}(q_1,q_2) = g\epsilon^{*d}_GT^d_{ij}e^{*{\mu}}_G
P_{\mu}(q_1,q_2)~.
\label{z7}
\en
Here $(T^a)_{ij} = -if^a_{ij}$ stands for the matrix elements of the
colour group generators in the adjoint representation
($f^a_{bc}$ being the group structure constants),
$\epsilon^d_G$ and $e^{\mu}_G$ are the colour wave
function and the polarization vector of the produced gluon with momentum
$k=q_1-q_2$.
The vector $P_{\mu}(q_1,q_2)$, given by
\eq
P^{\mu}(q_1,q_2) = -q^{\mu}_1-q^{\mu}_2+{p^{\mu}_A\over
{p_A\cdot k}}\left(t_1+\vec k^{\thinspace 2}_{\perp}\right)-{p^{\mu}_B\over
{p_B\cdot k}}
\left(t_2+\vec k^{\thinspace 2}_{\perp}\right)~,
\label{z8}
\en
has the property
\eq
k^{\mu}P_{\mu}(q_1,q_2) = 0~,
\label{z9}
\en
which makes transparent the gauge invariance of the amplitude (\ref{z5}).
\vskip.3cm
The program of the calculation of corrections to the LLA, presented in
ref.~\cite{LF},
includes, as a necessary step, the calculation of the radiative corrections
to the PPR and RRG vertices and to the gluon trajectory.
Corrections to the PPR vertices
were calculated in refs. [3-5]; they lead to a new (compared with the
feature found in the LLA) physical phenomenon - non conservation of the
helicity for
each of the colliding gluons in pure gluodynamics~\cite{FL}.
In QCD the situation can change: this effect indeed is absent for
the case of three massless quark flavours~\cite{FF}.
\vskip.3cm
Gluon loop corrections to the RRG vertex were calculated in ref.~\cite{FL}.
There a new effect also appears: the vertex, though it remains transverse,
cannot be expressed only in terms of the transverse vector
$P_{\mu}$ defined in eq.(\ref{z8}). In such a situation it becomes
very interesting to calculate the quark loop corrections to the RRG vertex.
This is all the more interesting because
quarks do not play any role in the LLA, therefore their role in the
perturbative pomeron formation could be understood only through the
corrections to the LLA.
\vskip.3cm
In this paper we calculate the quark loop contribution to the RRG vertex.
Section 2 is devoted to discuss the general structure
of the gluon production amplitude. Section 3 deals with
the calculation of the correction to the gluon production amplitude in the
framework of the dispersive approach. In section 4 the
quark contribution to the RRG vertex is obtained. Finally, in section
5 the spin structure of the quark correction to the RRG vertex
is analyzed.

\vskip 0.5cm

{\bf 2. General structure of the gluon production amplitude}

For reaching our goal, we use the dispersive approach based on the
analyticity and $t$-channel unitarity,
developed in refs. [3-5]. In order to obtain the corrections to the RRG
vertex one needs to calculate the corrections to the gluon
production amplitude and then compare this with its multi-Regge form,
corresponding to the contributions of the reggeized gluons in $t_1$ and $t_2$
channels. It is necessary to remind that the simple form (\ref{z5}) is
valid only in the LLA, where one does not make difference between $ln(s)$
and $ln(-s)$. Going beyond the LLA, from requirements of analyticity,
unitarity and crossing symmetry one arrives to the following general form
for the gluon production amplitude~\cite{Ba}:
\eqn
{\cal A}^{A'GB'}_{AB} = \frac{s}{4}\Gamma^{i_1}_{A'A}(t_1){1\over t_1}
\epsilon^{*d}_GT^d_{i_2i_1}{1\over t_2}\Gamma^{i_2}_{B'B}(t_2)
\enn
\eqn
\times \left\{\left[\left({s_1\over {\mu^2}}\right)^{{\omega_1}-
{\omega_2}}+\left({-s_1\over {\mu^2}}\right)^{{\omega_1}-{\omega_2}}
\right]\left[
\vbox to 22.9pt{}\left({s\over {\mu^2}}\right)^{\omega_2}+\left({-s\over
{\mu^2}}\right)^{\omega_2}\vbox to 22.9pt{}
\right]{\cal R}_G(t_1,t_2,\vec k_{\perp}^
{\thinspace 2})\right.
\enn
\eq
\left. +\left[\left({s_2\over {\mu^2}}\right)^{{\omega_2}-
{\omega_1}}+\left({-s_2\over {\mu^2}}\right)^{{\omega_2}-{\omega_1}}
\right]\left[\vbox to 22.9pt{}
\left({s\over {\mu^2}}\right)^{\omega_1}+\left({-s\over
{\mu^2}}\right)^{\omega_1}\vbox to 22.9pt{}
\right]{\cal L}_G(t_1,t_2,\vec k_{\perp}^
{\thinspace 2})\right\}~,
\label{z10}
\en
where
\eqn
\omega_i\equiv \omega(t_i)~,~~~~s_1=(p_{A'}+k)^2~,~~~~s_2=(p_{B'}+k)^2~,~~~~
\vec k_{\perp}^{\thinspace 2}=
{{s_1s_2}\over s}~.
\enn
The PPR vertices $\Gamma^i_{P'P}$ depend on the polarization of
particles $P$ and $P'$ and the
squared momentum transfer $t$. They are real for $t<0$. The Born
expression for these vertices is given in eq.(\ref{z6}); one loop
corrections were calculated in refs. [3-5].
\vskip.3cm
The RRG vertices ${\cal R}_G$ and ${\cal L}_G$ depend on the polarization of
the gluon, its transverse momentum $\vec k_{\perp}^{\thinspace 2}$ and
the squared transferred momenta $t_1$,
$t_2$. They are real in all channels where $t_{1,2} < 0$, $\vec k_{\perp}^
{\thinspace 2} > 0$. It is clear from eq.(\ref{z10}) that in each order of
perturbation theory the contribution of the sum ${\cal R}_G+{\cal L}_G$ is
leading while that of the difference ${\cal R}_G-{\cal L}_G$ is subleading.
In the LLA only the sum ${\cal R}_G+{\cal L}_G$ in the Born approximation
contributes; one has
\eq
{\cal R}^{(0)}_G+{\cal L}^{(0)}_G = -2ge^{*{\mu}}_GP_{\mu}
(q_1,q_2)~.
\label{z11}
\en
On the contrary, the difference ${\cal R}^{(0)}_G-{\cal L}^{(0)}_G$ at
the same order
g contributes to the amplitude (\ref{z10}) only as a radiative correction.
It can be obtained, together with one loop corrections of order $g^3$ to the
sum ${\cal R}^{(1)}_G+{\cal L}^{(1)}_G$, from the gluon production amplitude,
calculated with one loop accuracy assuming that the one loop corrections
$\Gamma^{(1)i}_{P'P}$ are known. In fact with such an accuracy from
eq.(\ref{z10}) we get
\eqn
{\cal A}^{A'GB'}_{AB}(one~loop) =
s{1\over t_1}\epsilon^{*d}_GT^d_{{i_2}{i_1}}{1\over t_2}
\left\{\vbox to 22.9pt{}-2ge^{*{\mu}}_
GP_{\mu}(q_1,q_2)\left[\vbox to 16.36pt{}
\Gamma^{(1)i_1}_{A'A}\Gamma^{(0)i_2}_{B'B}
+\Gamma^{(0)i_1}_{A'A}\Gamma^{(1)i_2}_{B'B}\right.\right.
\enn
\eqn
\left.\left.
+{1\over 4}\Gamma^{(0)i_1}_{A'A}\Gamma^{(0)i_2}_{B'B}\left((\omega^{(1)}_1+
\omega^{(1)}_2)ln\left({{s(-s)}\over{\mu^4}}\right)+(\omega^{(1)}_1-
\omega^{(1)}_2)ln\left({{s_1(-s_1)}\over {s_2(-s_2)}}\right)\right)\right]
+\Gamma^{(0)i_1}_{A'A}\Gamma^{(0)i_2}_{B'B}\right.
\enn
\eq
\left.\times \left[{{{\cal R}^{(0)}_G-{\cal L}^{(0)}_G}\over 4}(\omega^{(1)}_
1-\omega^{(1)}_2)ln\left({{s_1(-s_1)s_2(-s_2)}\over {s(-s)\mu^4}}\right)+
{\cal R}^{(1)}_G+{\cal L}^{(1)}_G\right]\right\}~.
\label{z12}
\en
We notice from eq.(\ref{z12}) that the difference
${\cal R}^{(0)}_G-{\cal L}^{(0)}_G$
contributes to the discontinuities of the gluon production
amplitude in $s_1$, $s_2$ and $s$ channels, therefore it can be found using
$s_i$-channel unitarity conditions~\cite{FKL,Ba}. As for the sum
${\cal R}^{(1)}_G+{\cal L}^{(1)}_G$, it cannot be defined by means of the
$s_i$-channel unitarity in the multi-Regge region. The $t$-channel approach
is indeed more suitable for this purpose. This sum has been calculated for
the case of pure gluodynamics~\cite{FL}. For the real case of QCD one needs
to add the quark contribution.
\vskip.3cm
It is possible to calculate the corrections to the RRG vertex by considering
the gluon production in various scattering processes: gluon-gluon (GG),
quark-quark (QQ) and quark-gluon (QG). Of course, we should obtain the same
vertex. In the approach based on $t_i$-channel unitarity relations it looks
very natural. One can consider all the three processes simultaneously.

\vskip 0.5cm

{\bf 3. Corrections connected with $t_1$-channel discontinuity}

Let us take into account the contribution of the amplitude discontinuity in
the $t_1$-channel. The discontinuity in the $t_2$-channel could be considered
analogously. The contribution is represented schematically in Fig. 2, where
particles $C$ and $C'$ are quarks, as we are interested in the quark-antiquark
intermediate state. In order to calculate this contribution, we need to
consider both amplitudes ${\cal A}^{A'C'}_{AC}$ and ${\cal A}^{CGB'}_{C'B}$.
On one hand, we must use an exact expression for the amplitude ${\cal A}^
{A'C'}_{AC}$, as particles $C$ and $C'$ are in the intermediate state and we
integrate over their momenta. On the other hand, since $s_2 = p_{B'}+k$
is fixed and large, we take the amplitude ${\cal A}^{CGB'}_{C'B}$ in
the quasi-multi-Regge kinematics~\cite{LF}, which means that the gluon $G$
is produced in the fragmentation region of the quark $C'$ . An expression
for the part of the amplitude with the gluon quantum numbers in $t_i$
channels can be written at once for both possible choices (quarks or gluons)
of particles $B$ and $B'$. It reads
\eqn
{\cal A}^{CGB'}_{C'B} = g^2T^{i_1}_{CC'}\bar u(p_{C})\left\{-\pnot_{B}\frac
{\pnot_{C'}-\knot+m_C}{(p_{C'}-k)^2-m^2_C}\enot^*_{G}\right.
\enn
\eqn
\left. +\enot^*_{G}\frac {\pnot_{C}+\knot+m_C}{(p_{C}+k)^2-m^2_C}
\pnot_{B}\right.\left. +\gamma^{\mu}{2\over{t_1}}
\left[\vbox to 13.09pt{}-(e^*_{G}\cdot(q_1+
q_2))p^{\mu}_B+s_2e^{*\mu}_{G}\right.\right.
\enn
\eq
\left.\left. +2(e^*_{G}\cdot p_B)\left(q^{\mu}_2-p^{\mu}_B{t_2\over s_2}
\right)\right]\vbox to 16,36pt{}
\right\}u(p_{C'})\epsilon^{*d}_GT^d_{i_1i_2}{1\over t_2}\Gamma^{(0)i_
2}_{B'B}~.
\label{z13}
\en
\vskip.3cm
As for the amplitude ${\cal A}^{A'C'}_{AC}$, it is very profitable here to
decompose it into two parts [3-5] which are schematically shown in
Fig. 3:
\eq
{\cal A}^{A'C'}_{AC}={\cal A}^{A'C'}_{AC}(as)+{\cal A}^{A'C'}_{AC}(na)~.
\label{z14}
\en
The first term in the r.h.s. of eq.(\ref{z14}) contains the contribution
which is proportional to $s_A=(p_A+p_C)^2$ in the asymptotic region
$s_A \gg {|t_1|}$, while the second one cannot include contributions
increasing with $s_A$ in this region. The first term can be written in the
same form for any kind (quarks or gluons) of particles $A$ and $A'$
(cf.~\cite{FF,FFQ}):
\eq
{\cal A}^{A'C'}_{AC}(as)=2g\Gamma^{(0)i}_{A'A}{1\over t_1}T^i_{C'C}
\bar u(p_{C'})\pnot_{A}u(p_{C})~.
\label{z15}
\en
The explicit form of the second term is not quoted because we will not need
it in our calculations, as it will be discussed in the next section.
\vskip.3cm
According to the decomposition (\ref{z14}), we split the contribution
represented in Fig. 2 into two terms shown in Fig.s 4$(a)$ and 4$(b)$.
Let us firstly consider the contribution of Fig. 4$(a)$.
Instead of the discontinuity, we calculate the contribution to the
amplitude ${\cal A}^{A'GB'}_{AB}$ itself, substituting the full denominator
$i(p^2-m^2+i\varepsilon)^{-1}$ of the Feynman propagator for an intermediate
particle with momentum $p$ in place of $2\pi\delta(p^2-m^2)$ in the
expression for the discontinuity.
Using the expressions (\ref{z13}) and (\ref{z15}) for the r.h.s. and l.h.s.
amplitudes respectively, the contribution takes the form
\eqn
{\cal A}^{(a)A'GB'}_{AB} = 2g^3s\Gamma^{(0)i_1}_{A'A}{1\over t_1}
\epsilon^{*d}_GT^d_{i_1i_2}{1\over t_2}\Gamma^{(0)i_2}_{B'B}
{{p^{\mu}_Ap^{\nu}_B}\over s}e^{*\rho}_G
\enn
\eq
\times \sum_{f}\left[{2\over t_1}{\cal P}^{f}_{\mu\nu}(q_1)\left(P
(q_1,q_2)-2p_A{t_1\over s_1}\right)_{\rho}+{\cal V}^{f}_{\mu\nu\rho}(q_1,q_2)
\right]~,
\label{z16}
\en
where the summation is performed over quark flavours and the vertex vector
$P_{\mu}(q_1,q_2)$ is given by eq.(\ref{z8}). The tensor
\eqn
{\cal P}^{f}_{\mu\nu}(q_1)= {i\over 2}\int{d^Dp\over{({2\pi})^D}}{
tr\left[\gamma^{\mu}(\pnot+m_f)\gamma^{\nu}(\pnot+\qnot+m_f)\right]
\over{(p^2-m_f^2+i\varepsilon)((p+q)^2-m_f^2+i\varepsilon)}} =
(g_{\mu \nu}q^2-q_{\mu}q_{\nu}){\cal P}^f(q^2)~,
\enn
\eq
{\cal P}^f(q^2)=-4{\Gamma\left(2-{D\over 2}\right)\over {{(4\pi)}^{D\over 2}}}
\int^1_0{{dx}x(1-x)\over {\left(m_f^2-q^2x(1-x)\right)}^{2-{D\over 2}}}
\label{z17}
\en
is the well known fermion contribution to the gluon polarization operator,
and the tensor
\eqn
{\cal V}^{f}_{{\mu}{\nu}{\rho}}(q_1,q_2) =
\enn
\eq
-i\int{d^Dp\over{({2\pi})^{D}}}{
tr\left[\gamma_{\mu}(\pnot+m_f)\gamma_\nu(\pnot+\qnot_2+m_f)\gamma_{\rho}
(\pnot+\qnot_1+m_f)\right]\over(p^2-m_f^2+i\epsilon)((p+q_1)^2-m_f^2+i
\epsilon)((p+q_2)^2-m_f^2+i\epsilon)}
\label{z18}
\en
represents the quark contribution to the triple gluon vertex. In deriving
eq.(\ref{z16}) we used the symmetry of the first two terms of
eq.(\ref{z13}) under the
change $p_C\leftrightarrow \-p_{C'}$ to reduce their contribution to the
term ${\cal V}^{f}_{{\mu}{\nu}{\rho}}$ in eq.(\ref{z16}); on the other hand,
through the usual trick for $g^{\mu\nu}$, we approximated $\gamma^{\mu}$,
that appears in eq.(\ref{z13}), with
\eq
\gamma^{\mu}=\gamma_{\nu}g^{\mu\nu}\approx \gamma_{\nu}{{2p^{\mu}_Ap^
{\nu}_B}\over s}~.
\label{z19}
\en
Subsequently, performing the Feynman parametrization and the integration
over momentum $p$ and using also the property
\eqn
e^*_G\cdot k=e^*_G\cdot (q_1-q_2)=0
\enn
we get, in the multi-Regge asymptotic region, the following non vanishing
part of the convolution of the tensor
${{p^{\mu}_Ap^{\nu}_B}\over s}e^{*\rho}_G$
with the tensor ${\cal V}^{f}_{{\mu}{\nu}{\rho}}$ shown in eq.(\ref{z18}):
\eqn
{ { p^{\mu}_Ap^{\nu}_B } \over s } e^{*\rho}_G {\cal V}^{f}_{{\mu}{\nu}
{\rho} }(q_1,q_2)={{\Gamma\left(3-{D\over 2}\right)}\over {{(4\pi)}^
{D\over 2}}}\int^1_0 \int^1_0{{dx_1dx_2\theta(1-x_1-x_2)}\over {\left[R^{f}
(t_1,t_2)\right]^{3-\frac{D}{2}}}}
\enn
\eqn
\times \left\{\left[m^2_f+\left({2-D\over {D-4}}\right)R^{f}(t_1,t_2)\right]
\left[\vbox to 13.09pt{}-\left(e^*_G\cdot (q_1+q_2)\right)(2-x_1-x_2)
\right.\right.
\enn
\eqn
\left.\left. +2{s_2\over s}(e^*_G\cdot p_A)(1+x_1)-2{s_1\over s}(e^*_G\cdot
p_B)(1+x_2)\right]\right.
\enn
\eqn
\left. +(1-x_1-x_2)\left[\left(e^*_G\cdot (q_1+q_2)\right)\left(R^{f}(t_1,
t_2)-m^2_f+{{2s_1s_2}\over s}x_1x_2\right)\right.\right.
\enn
\eqn
\left.\left. +2{s_2\over s}(e^*_G\cdot p_A)(t_1x^2_1+t_2x_2(1+x_1))
\right.\right.
\enn
\eq
\left.\left. -2{s_1\over s}(e^*_G\cdot p_B)(t_2x^2_2+t_1x_1(1+x_2))\right]
\right\}~,
\label{z20}
\en
where
\eq
R^{f}(t_1,t_2)=m^2_f-(1-x_1-x_2)(t_1x_1+t_2x_2)~.
\label{z21}
\en
\vskip.3cm
Let us now turn to the contribution represented in Fig. 4$(b)$. It is
expressed in terms of the product
${\cal A}^{A'C'}_{AC}(na)\cdot {\cal A}^{CGB'}_{C'B}$. Since the
non-asymptotic part ${\cal A}^{A'C'}_{AC}(na)$ does not contain terms of
order $s_A$ for large values of this invariant, the essential region of
integration over $p_C$ in this case is
\eqn
s_A\sim (p_A-p_{C'})^2\sim p^2_C\sim p^2_{C'}\sim q^2_i~,~~~i=1,2~,
\enn
\eq
(k+p_C)^2\sim (k-p_{C'})^2\sim (p_{A'}+k)^2=s_1 ~.
\label{z22}
\en
This implies that, in order to calculate the contribution of Fig. 4$(b)$,
the amplitude ${\cal A}^{CGB'}_{C'B}$ may be taken in the multi-Regge
asymptotic region. Moreover, due to the relations
\eq
{{e^*_G\cdot p_C}\over {k\cdot p_C}}\approx {{e^*_G\cdot p_A}\over
{k\cdot p_A}}~,~~~~~~~~~~~{{k\cdot p_C}\over {p_B\cdot p_C}}\approx
{{k\cdot p_A}\over {p_B\cdot p_A}}~,
\label{z23}
\en
valid in the region (\ref{z22}), the amplitude can be written as
\eq
{\cal A}^{CGB'}_{C'B}=2gT^{i_1}_{CC'}\bar u(p_{C})\pnot_{B}u(p_{C'})
{1\over t_1}\gamma^G_{i_1i_2}(q_1,q_2){1\over t_2}\Gamma^{(0)i_2}_{B'B}~,
\label{z24}
\en
where the vertex $\gamma^G_{i_1i_2}(q_1,q_2)$ is given by eqs.(\ref{z7})
and (\ref{z8}). It means that the amplitude (\ref{z24}) differs from the
asymptotic part ${\cal A}^{CB'}_{C'B}(as)$ of the elastic scattering
amplitude only for factors which do not depend on the momenta
$p_C$ and $p_{C'}$ of the intermediate particles. Consequently, the
contribution of Fig. 4$(b)$ can be calculated in the
same way as the corresponding contribution to the elastic scattering
amplitude ${\cal A}^{A'B'}_{AB}$ coming from the product ${\cal A}^{A'C'}_
{AC}(na)\cdot {\cal A}^{CB'}_{C'B}(as)$ (see refs.~\cite{FF,FFQ}).
As a matter of fact, it is not necessary to calculate this contribution
at all, since it is totally absorbed by the corrections to the PPR vertices
$\Gamma^{(1)i}_{P'P}$. Let us stress that one-loop corrections to the gluon
production amplitude include the corrections to these vertices, as one may
observe in eq.(\ref{z12}).

\vskip 0.5cm

{\bf 4. Quark contribution to the reggeon-reggeon-gluon vertex}

In order to obtain the corrections to the RRG vertex one could calculate the
corrections to the gluon production amplitude and subtract those ones coming
from
the PPR vertices. However, it is more preferable to identify and subtract
these last corrections without calculating them. This indeed can be easily
done by comparing the corrections
to the elastic scattering amplitude with those ones to the gluon production
amplitude. In the case of the elastic scattering the corrections connected
with the quark-antiquark
$t$-channel intermediate state are expressed in terms of the corrections to
the PPR vertices $\Gamma^{(1)i}_{A'A}$ and $\Gamma^{(1)i}_{B'B}$. They are
found
by using the $t$-channel unitarity and by performing the decomposition
(\ref{z14}) for
${\cal A}^{A'C'}_{AC}$ and the analogous one for ${\cal A}^{CB'}_{C'B}$.
The contribution coming from ${\cal A}^{A'C'}_{AC}(na)\cdot {\cal A}^{CB'}_
{C'B}(as)$ represents a part of the corrections to the amplitude connected
with a piece of $\Gamma^{(1)i}_{A'A}$. In the case of the gluon production,
because of the factorization property
of ${\cal A}^{CB'}_{C'B}(as)$ (cf. eq.(\ref{z15})) and the analogous
property of ${\cal A}^{CGB'}_{C'B}$ (cf. eq.(\ref{z24})), the correction
due to ${\cal A}^{A'C'}_{AC}(na)\cdot {\cal A}^{CGB'}_{C'B}$ is connected
with the same piece of $\Gamma^{(1)i}_{A'A}$.
Therefore we may exclude this piece from the corrections to the
gluon production
amplitude and thus we do not need to calculate it. After that, we have to
consider only the piece of $\Gamma^{(1)i}_{A'A}$ defined by
${\cal A}^{A'C'}_{AC}(as)\cdot {\cal A}^{CB'}_{C'B}(as)$ and subtract from
the corrections to the gluon production amplitude only the part connected
with this piece. We know from refs.~\cite{FF,FFQ} that
\eq
\Gamma^{(1)i}_{A'A}(as\cdot as)=
g^2\Gamma^{(0)i}_{A'A}{1\over t_1}{{p^{\mu}_Ap^
{\nu}_B}\over s}\sum_{f}{\cal P}^{f}_{\mu\nu}(q_1)~.
\label{z25}
\en
Using eq.(\ref{z25}) and comparing eqs.(\ref{z12}) and (\ref{z16}) we
conclude that, in order to avoid taking into consideration the corrections
to $\Gamma^{i}_{A'A}$, we only need to divide by a factor 2 the term with
$P_{\mu}(q_1,q_2)$ in eq.(\ref{z16}).
\vskip.3cm
Consequently, the part of the corrections to the the RRG vertex connected
with the quark-antiquark intermediate state in the $t_1$-channel is given by
\eqn
({\cal R}^{(1)}_G+{\cal L}^{(1)}_G)_{t_1}=
\enn
\eq
-2g^3{{p^{\mu}_Ap^{\nu}_B}\over
s}e^{*\rho}_G\sum_{f}\left[{1\over t_1}{\cal P}^{f}_{\mu\nu}(q_1)\left(P
(q_1,q_2)-4p_A{t_1\over s_1}\right)_{\rho}+{\cal V}^{f}_{\mu\nu\rho}(q_1,
q_2)\right]~,
\label{z26}
\en
where $P_{\mu}(q_1,q_2)$, ${\cal P}^{f}_{\mu\nu}(q_1)$ and ${\cal V}^{f}_
{\mu\nu\rho}(q_1,q_2)$ are given by eqs.(\ref{z8}), (\ref{z17}) and
(\ref{z18}) respectively.
\vskip.3cm
The sum in eq.(\ref{z26}) has a correct discontinuity in the $t_1$-channel.
It has also a $t_2$-channel discontinuity, but this is correct only for
terms which have both $t_1$ and $t_2$-channel discontinuities. An
expression for the sum ${\cal R}^{(1)}_G+{\cal L}^{(1)}_G$ with correct
discontinuities
in both channels can be easily yielded by symmetry considerations. Let us
notice that, due to the colour structure of the amplitude (\ref{z10}) and
the momentum flow (see Fig. 1), the sum ${\cal R}^{(1)}_G+{\cal L}^{(1)}_G$
must change sign under the substitution
\eq
q_1\leftrightarrow -q_2~,~~~~~~~~~~~p_A\leftrightarrow p_B~,
\label{z27}
\en
as the vector $P_{\mu}(q_1,q_2)$ does. The function ${ { p^{\mu}_Ap^
{\nu}_B } \over s } e^{*\rho}_G {\cal V}^{f}_{{\mu}{\nu}
{\rho} }(q_1,q_2)$, given in eq.(\ref{z20}), has such a property, so we
obtain correct discontinuities in both $t_i$ channels if we add the term
\eq
{1\over t_2}{\cal P}^{f}_{\mu\nu}(q_2)\left(P(q_1,q_2)+4p_B{t_2\over
s_2}\right)_{\rho}
\label{z28}
\en
into the square brackets of eq.(\ref{z26}).
\vskip.3cm
Contrary to the massless case, where correct analytical properties together
with the unitarity requirement in $t_i$ channels determine the amplitude
in an unambiguous way~\cite{FL}, in the massive quark case we can add an
expression which is
equal to the Born amplitude with some constant coefficient~\cite{FFQ}. If
we want to use a customary renormalization scheme, this
coefficient should be
determined by Feynman diagrams. This means that we need to add the
contribution of Feynman diagrams with a quark loop inserted in the produced
gluon line. This contribution is equal to the Born amplitude with the
coefficient
\eq
{g^2\over 2}\sum_{f}{\cal P}^{f}(0) =
-{g^2\over 3}{\Gamma\left(2-{D\over 2}\right)\over {{(4\pi)
}^{D\over 2}}}\sum_{f}(m^2_f)^{{D\over 2}-2}~.
\label{z29}
\en
The last equality can be easily derived from the second of eqs.(\ref{z17}).
\vskip.3cm
Finally, we find that the quark contribution to the one loop corrections to
the RRG vertex is given by
\eqn
{\cal R}^{(1)}_G+{\cal L}^{(1)}_G=
-g^3e^{*\rho}_G\sum_{f}\left[\vbox to 16.36pt{}P_{\rho}(q_1,q_2)
\left({\cal P}^{f}
(t_1)+{\cal P}^{f}(t_2)+{\cal P}^{f}(0)\right)\right.
\enn
\eq
\left. -4{{p_{A\rho}t_1}\over s_1}{\cal P}^{f}(t_1)+4{{p_{B\rho}t_2}\over
s_2}{\cal P}^{f}(t_2)+2{{p^{\mu}_Ap^{\nu}_B}\over s}{\cal V}^{f}_{\mu\nu
\rho}(q_1,q_2)\right]~.
\label{z30}
\en
By integrating eq.(\ref{z20}) over one of the two $x$ variables keeping fixed
their sum, the quantity ${\cal R}^{(1)}_G+{\cal L}^{(1)}_G$ can be written
in a manifestly gauge invariant form:
\eqn
{\cal R}^{(1)}_G+{\cal L}^{(1)}_G=
\enn
\eq
e^{*\rho}_G\sum_{f}\left[P_{\rho}(q_1,q_2)(a^{f}_1+a^{f}_2)+
\left({p_A\over s_1}-{p_B\over s_2}\right)_
{\rho}(-(t_1+t_2+2\vec k^{\thinspace 2}_{\perp})a^{f}_2+a^{f}_3)\right]~.
\label{z31}
\en
The coefficients $a_i$ in eq.(\ref{z31}) depend on $t_i$ and
$\vec k^{\thinspace 2}_{\perp}$ and read
\eqn
a^{f}_1=g^3\left[{{t_1+t_2}\over {t_1-t_2}}\left({\cal P}^{f}(t_1)-{\cal P}^
{f}(t_2)\right)-{\cal P}^{f}(0)\right]~,
\enn
\eqn
a^{f}_2={8g^3\over {D-2}}{\Gamma\left(3-{D\over 2}\right)\over {{(4\pi)}^{D
\over 2}}}{\vec k^{\thinspace 2}_{\perp}\over {(t_1-t_2)^3}}
\enn
\eqn
\times \int^1_0{dz\over z^2}\left[{{2R_1R_2}\over {D-4}}\left(R^{{D\over
2}-2}_1-R^{{D\over 2}-2}_2\right)-{2\over D}\left(R^{D\over 2}_1-R^{D\over
2}_2\right)\right]~,
\enn
\eqn
a^{f}_3=g^3{\Gamma\left(2-{D\over 2}\right)\over {{(4\pi)}^{D\over 2}}}
{4\over {t_1-t_2}}\int^1_0dz
\left[\vbox to 22.9pt{}4t_1t_2z(1-z)\left(R^{{D\over 2}-2}_1-
R^{{D\over 2}-2}_2\right)\right.
\enn
\eq
\left. +{2\vec k^{\thinspace 2}_{\perp}\over {D-2}}\left(3-{1\over {2z
(1-z)}}\right)\left(R^{{D\over 2}-1}_1-R^{{D\over 2}-1}_2\right)\right]~,
\label{z32}
\en
where
\eq
R_i=m^2_f-z(1-z)t_i~.
\label{z33}
\en
Let us notice that the coefficients $a^{f}_2$ and $a^{f}_3$ do not
contain ultraviolet (as well as infrared) singularities. Only the term
${\cal P}^{f}(0)$ has such singularities. Evidently, poles at $t_1=t_2$ in
formulae (\ref{z32}) for $a^{f}_i$ are fictitious. One can verify that they
cancel and the coefficients have only logarithmic singularities in $t_1$ and
$t_2$.
\vskip.3cm
In the massless case the integrals in eqs.(\ref{z32}) can be calculated
for arbitrary $D$ yielding
\eqn
a^{f}_1|_{m_f=0}= -4g^3{\Gamma\left(2-{D\over 2}\right)\over {{(4\pi)}^
{D\over 2}}}{\Gamma^2\left({D\over 2}\right)\over{\Gamma(D)}}{{t_1+t_2}
\over {t_1-t_2}}\left((-t_1)^{{D\over 2}-2}-(-t_2)^{{D\over 2}-2}\right)~,
\enn
\eqn
a^{f}_2|_{m_f=0}= -4g^3{\Gamma\left(2-{D\over 2}\right)\over {{(4\pi)}^
{D\over 2}}}{\Gamma^2\left({D\over 2}-1\right)\over{\Gamma(D)}}{\vec k^
{\thinspace 2}_{\perp}\over {(t_1-t_2)^3}}
\enn
\eqn
\times \left[{D\over 2}t_1t_2\left((-t_1)^{{D\over 2}-2}-(-t_2)^{{D\over 2}
-2}\right)+(2-{D\over 2})\left((-t_1)^{D\over 2}-(-t_2)^{D\over 2}\right)
\right]~,
\enn
\eqn
a^{f}_3|_{m_f=0}=4g^3{\Gamma\left(2-{D\over 2}\right)\over {{(4\pi)}^
{D\over 2}}}{\Gamma^2\left({D\over 2}-1\right)\over{\Gamma(D)}}
\enn
\eqn
\times {1\over {t_1-t_2}}
\left[\vbox to 13.09pt{}(D-2)^2t_1t_2\left((-t_1)^{{D\over 2}-2}-
(-t_2)^{{D\over 2}-2}\right)\right.
\enn
\eq
\left. -\left(2-{D\over 2}\right)\vec k^{\thinspace 2}_{\perp}\left((-t_1)^
{{D\over 2}-1}-(-t_2)^{{D\over 2}-1}\right)\right]~.
\label{z34}
\en
It is worth noticing that these coefficients have not any singularity at
$D=4$. However, the coefficient $a^f_1$
must have an ultraviolet singularity in order to cancel the corresponding
one in the sum ${\cal R}_G+{\cal L}_G$ when this quantity is expressed in
terms of the
renormalized coupling constant. If $g_{\mu}$ is such a constant in the
${\overline {MS}}$ scheme at the renormalization point $\mu$, then one has
\eq
g=g_{\mu}\mu^{2-{D\over 2}}\left\{1+\left({11\over 3}N-{2\over 3}n_f\right)
{g^2_{\mu}\over {(4\pi)^2}}\left[{1\over {D-4}}-{1\over 2}(ln(4\pi)-\gamma)
\right]+\cdots\right\}~,
\label{z35}
\en
where $\gamma$ is the Euler constant and the term with the quark flavour
number $n_f$ yields the ``corresponding" quark induced singularity. The gluon
induced singularity, originating from the term with $N$, is absorbed by the
gluon contribution to the RRG vertex~\cite{FL}. At first sight the absence
of the singularity in eq.(\ref{z34}) could disturb, but one  should realize
that the term  with the ultraviolet singularity is cancelled there by the
one with the infrared singularity~\cite{HV} which arises in
${\cal P}^{f}(0)$ at $m_f=0$.
\vskip.3cm
In the massive quark case for arbitrary $D$ the integrals in eq.(\ref{z32})
cannot be expressed in terms of elementary functions  but it is possible for
$D=4$.
After performing the charge renormalization (\ref{z35}), taking into account
that the gluon induced terms in eq.(\ref{z35}) are absorbed by the
gluon contribution to the RRG vertex,
the sum ${\cal R}^{(1)}_G+{\cal L}^{(1)}_G$ of order $g^3_{\mu}$ to
which we arrive is given by eq.(\ref{z31}) with the following coefficients:
\eqn
a^{f}_{1R}|_{D=4}={2\over 3}{g^3_{\mu}\over {(4\pi)^2}}\left\{{{4m^2_f(t_1
+t_2)}\over {t_1t_2}}+{{t_1+t_2}\over {t_1-t_2}}\left[\left(2+{{4m^2_f}
\over t_1}\right){L_1\over {\beta_1}}\right.\right.
\enn
\eqn
\left.\left. -\left(2+{{4m^2_f}\over t_2}\right){L_2\over {\beta_2}}\right]-
ln\left({m^2_f\over {{\mu}^2}}\right)\right\}~,
\enn
\eqn
a^{f}_2|_{D=4}=4{g^3_{\mu}\over {(4\pi)^2}}{{\vec k^{\thinspace 2}_{\perp}}
\over {(t_1-t_2)^3}}\left[\vbox to 16.36pt{}-2m^2_f(t_1+t_2)(L^2_1-L^2_2)+
\left(\vbox to 13.09pt{}4m^2_f(t_1+t_2)
\right.\right.
\enn
\eqn
\left.\left. +{2\over 3}t_1t_2\right)\left({L_1\over {\beta_1}}-{L_2\over
{\beta_2}}\right)-{{8m^2_f}\over 3}\left({t_2{L_1}\over {\beta_1}}-{{t_1L_2}
\over {\beta_2}}\right)-{{(t_1-t_2)}\over 6}(t_1+t_2+16m^2_f)\right]~,
\enn
\eqn
a^{f}_3|_{D=4}=4{g^3_{\mu}\over {(4\pi)^2}}\left\{{{\vec k^{\thinspace 2}_
{\perp}}\over {t_1-t_2}}\left[2m^2_f(L^2_1-L^2_2)-4m^2_f\left({L_1
\over {\beta_1}}-{L_2\over {\beta_2}}\right)+{{t_1-t_2}\over 6}\right]\right.
\enn
\eq
\left. -{4\over {3(t_1-t_2)}}\left({{t_2(2m^2_f+t_1)}\over {\beta_1}}L_1-
{{t_1(2m^2_f+t_2)}\over {\beta_2}}L_2\right)-{8\over 3}m^2_f
\vbox to 22.9pt{}\right\}~,
\label{z36}
\en
where the index $R$ in $a^{f}_1$ denotes that the renormalization was
performed and
\eq
\beta=\sqrt{{-t\over {4m^2_f-t}}}~,~~~~~~L={1\over 2}ln\left({{1+\beta}\over
{1-\beta}}\right)=ln\left(\sqrt{1-\frac{t}{4m^2_f}}+\sqrt{\frac {-t}{4m^2_f}}
\right).
\label{z37}
\en
\vskip.3cm
Finally, let us consider the massless quark case in the physical space-time
dimension $D=4$. We may approach this case moving either from
eq.(\ref{z36})
or from eq.(\ref{z34}) taking into account the charge renormalization
(\ref{z35}) in
the Born term (\ref{z11}). These two ways are characterized by different
regularizations of the infrared singularity which arises in
${\cal P}^f(0)$
at $m_f=0$. Following the second approach we get
\eqn
{\mu}^{\frac{D}{2}-2}\left({\cal R}^{(1)}_G+{\cal L}^{(1)}_G\right)|_{{m_f=0}
\atop {D\rightarrow 4}}=
\enn
\eqn
4{g^3_{\mu}\over {(4\pi)^2}}n_f
\left\{\vbox to 22.9pt{}\frac{1}{3}e^{*\rho}_GP_
{\rho}(q_1,q_2)
\left[\vbox to 22.9pt{}\frac{1}{D-4}-\frac{1}{2}(ln(4\pi)-\gamma)+
\left(\frac{t_1+t_2}{2(t_1-t_2)}\right.\right.\right.
\enn
\eqn
\left.\left.\left. +\frac{\vec k^{\thinspace 2}_{\perp}t_1t_2}{(t_1-t_2)^3}
\right)ln\left(\frac{t_1}{t_2}\right)-\frac{\vec k^{\thinspace 2}_{\perp}
(t_1+t_2)}{2(t_1-t_2)^2}\right]+e^{*\rho}_G\left(\frac{p_A}{s_1}-\frac{p_B}
{s_2}\right)_{\rho}\right.
\enn
\eqn
\left.\times \left[\frac{t_1t_2}{3(t_1-t_2)}\left(-2-\frac{\vec k^
{\thinspace 2}_{\perp}(t_1+t_2+2\vec k^{\thinspace 2}_{\perp})}{(t_1-t_2)^2}
\right)ln\left(\frac{t_1}{t_2}\right)\right.\right.
\enn
\eq
\left.\left. +\frac{\vec k^{\thinspace 2}_{\perp}}{6}\left(1+\frac{(t_1+t_2)
(t_1+t_2+2\vec k^{\thinspace 2}_{\perp})}{(t_1-t_2)^2}\right)\right]\right\}~.
\label{z38}
\en
If the infrared singularity was regularized by mass $m$, in eq.(\ref{z38})
we would have $ln\left(\frac{\mu}{m}\right)$ instead of
$\frac{1}{D-4}-\frac{1}{2}(ln(4\pi)-\gamma)$.

\vskip 0.5cm

{\bf 5. Conclusion}

Let us discuss the results we obtained. We  already mentioned that poles at
$t_1=t_2$ in eqs.(\ref{z32}), (\ref{z34}), (\ref{z36}) and (\ref{z38}) are
fictitious and the vertex has only logarithmic singularities in $t_1$ and
$t_2$.
As for the dependence on $\vec k^{\thinspace 2}_{\perp}$, it is only
polynomial for the quark contribution considered here. There are some
a priori restrictions on the coefficients with which the spin structures
$e^{*\mu}_GP_{\mu}(q_1,q_2)$ and $e^{*\mu}_G\left(\frac{p_A}{s_1}
-\frac{p_B}{s_2}\right)_{\mu}$ enter the vertex. Such restrictions are
connected with the structure of QCD in the infrared region, where only
logarithmic singularities are permitted. One can easily verify from
eq.(\ref{z8}) that, when $|\vec q_{1\perp}|$ or $|\vec q_{2\perp}|$ tends to
zero, the vector $P_{\mu}(q_1,q_2)$ becomes proportional to $k_{\mu}$,
which in turn means that $e^{*\mu}_GP_{\mu}(q_1,q_2)$ tends to zero.
This last result can be proved summing up over the physical polarization
states:
\eq
\sum_{\lambda_G=1,2}|e^{*\mu}_GP_{\mu}(q_1,q_2)|^2=-\left(P
(q_1,q_2)\right)^2=\frac{4\vec q_{1\perp}^{\thinspace 2}\vec q_{2\perp}^
{\thinspace 2}}
{\vec k^{\thinspace 2}_{\perp}}~.
\label{z39}
\en
\vskip.3cm
This property of $P_{\mu}(q_1,q_2)$ guarantees a correct infrared behaviour
of the vertex and determines, in the Born approximation, the spin structure
of the RRG vertex (\ref{z7}) in a unique way. The matter is that the other
spin structure does not vanish at $\vec q_{1\perp}$ or $\vec q_{2\perp}$
equal to zero:
\eq
\sum_{\lambda_G=1,2}|e^{*\mu}_G\left(\frac{p_A}{s_1}-\frac{p_B}{s_2}\right)
_{\mu}| ^2=\frac{1}{\vec k^{\thinspace 2}_{\perp}}~.
\label{z40}
\en
It means that this spin structure can enter the vertex only with a
coefficient which turns to zero at $\vec q_{1\perp}$ or $\vec q_{2\perp}$
equal to zero. In the Born approximation a coefficient with such properties
cannot be found, as it could only exhibit $t_1$ and $t_2$ singularities at
infinity. This spin structure may instead appear in the corrections as it
really does, as well as in the gluon contribution to the RRG vertex
{}~\cite{FL}. On the contrary, for a coefficient of $e^{*\mu}_GP_{\mu}$ we
only require the absence of power singularities at small $\vec q_{1\perp}$
or $\vec q_{2\perp}$. One might check that these requirements are fulfilled
by inspecting eqs.(\ref{z31}), (\ref{z32}), (\ref{z34}), (\ref{z36}) and
(\ref{z38}).
\vskip.3cm
Comparing the coefficient of the new spin structure in eq.(\ref{z38})
with the corresponding one in the gluon contribution (see eq.(86) in the
second paper of ref.~\cite{FL}), one observes a striking similarity of both
coefficients. Furthermore, for $n_f=N$ we find almost full cancellation of
the two contributions to the new spin structure, apart from one term with
the coefficient $-\frac{2}{3}n_f$ for the quark case and $\frac{11}{3}N$
for the gluon case. Unfortunately, we have not an explanation for it.

\vskip 1.5cm
\underline {Acknowledgement}: One of us (V.F.) thanks the Dipartmento di
Fisica dell'Universit\`a della Calabria and the Istituto Nazionale di Fisica
Nucleare - Gruppo collegato di Cosenza for their warm hospitality while this
work was done.

\newpage

\centerline{\bf Figure Captions}
\vskip .3 cm
\begin{description}

\item{Fig. 1:}
Diagram for the multiple gluon production.
\item{Fig. 2:}
$t_1$-channel discontinuity of the gluon production amplitude.
\item{Fig. 3:}
Decomposition of the elastic scattering amplitude in its asymptotic and non
asymptotic parts.
\item{Fig. 4$(a)$:}
Contribution to $t_1$-channel discontinuity from the product of the
asymptotic part of the elastic amplitude with the gluon production
amplitude.
\item{Fig. 4$(b)$:}
Contribution to $t_1$-channel discontinuity from the product of the
non-asymptotic part of the elastic amplitude with the gluon production
amplitude.
\end{description}

\end{document}